\documentclass[
12pt,preprint,preprintnumbers,nofootinbib,groupedaddress,superscriptaddress,amsmath,amssymb]{revtex4}
\usepackage
{graphicx}
\usepackage{dcolumn}
\usepackage{bm}
\usepackage{amssymb}
\usepackage{amsmath}
\usepackage{epsfig}    
\usepackage
{color}
\usepackage{slashed}
\usepackage{hhline}

\def\be{\begin{equation}}
\def\ee{\end{equation}}
\newcommand{\bea}{\begin{eqnarray}}
\newcommand{\eea}{\end{eqnarray}}
\newcommand{\nn}{\nonumber}

\begin{document}


\title{Neutrino masses, dark matter and leptogenesis with  $U(1)_{B-L}$ gauge symmetry}
%

\author{Chao-Qiang Geng}
\email{geng@phys.nthu.edu.tw}
\affiliation{School of Physics and Information Engineering, Shanxi Normal University, Linfen 041004, China}
\affiliation{Physics Division, National Center for Theoretical Sciences, Hsinchu, Taiwan 300}
\affiliation{Department of Physics, National Tsing Hua University, Hsinchu, Taiwan 300}

\author{Hiroshi Okada}
\email{macokada3hiroshi@cts.nthu.edu.tw}
\affiliation{Physics Division, National Center for Theoretical Sciences, Hsinchu, Taiwan 300}

\date{\today}

\begin{abstract}
We  propose a model with an $U(1)_{B-L}$ gauge symmetry, in which  small neutrino masses,
dark matter and the matter-antimatter asymmetry in the Universe can be simultaneously explained.
In particular, the neutrino masses are generated radiatively, while the matter-antimatter asymmetry is led by 
the leptogenesis mechanism,  at TeV scale.
We also explore allowed regions of the model parameters and discuss some phenomenological effects, including
 lepton flavor violating processes.   
\end{abstract}
\maketitle
\newpage

\section{Introduction}
Radiative neutrino mass generation is one of the most promising candidates to naturally explain the small mass scales of active neutrinos.
Some models to realize this type of the approach  can also accommodate dark matter (DM).
Normally, when one considers the DM candidate in a theory, an additional symmetry 
such as $Z_2$ is imposed  in order to stabilize it.
The representative model has been shown in ref.~\cite{Ma:2006km}. 
In recent years, a lot of applications have been presented in the literature.
In particular, a model in ref.~\cite{Chen:2009gd} with introducing $Z_2\times Z_2$ symmetry and two inert isospin doublet bosons was proposed to understand  the cosmic ray  anomaly~\cite{CosmicRayAnomaly} by a decaying fermionic DM. 
In this model, the Baryon Asymmetry of the Universe (BAU) is understood via the leptogenesis mechanism within 
the TeV scale.

Our paper extends the study  in ref.~\cite{Chen:2009gd}  by having a gauged $U(1)_{B-L}$ symmetry instead of $Z_2\times Z_2$,
in which three right-handed neutrinos are naturally introduced as a usual model with the $U(1)_{B-L}$ symmetry.
However, their charge assignments are taken with a very unique manner, $i.e.$ $-4,-4$ and $5$ for three right-handed neutrinos~\cite{Montero:2007cd, Patra:2016ofq}, respectively.\footnote{Several applications along this ideas can be found in refs.~\cite{Nomura:2017vzp, Nomura:2017jxb, Nanda:2017bmi, DeRomeri:2017oxa}} 
It suggests that the first two right-handed neutrinos can contribute to the active neutrinos with masses,
while the third right-handed one can be a good DM candidate, even though two kinds of new bosons with nonzero $B-L$ charges have to be added to give the masses of right-handed neutrinos. As a result, the stabilized symmetry of DM ($Z_2$) is induced as a remnant symmetry after the spontaneous symmetry breaking (SSB) of $U(1)_{B-L}$.\footnote{{A $U(1)_{B-L}$ gauge symmetry is sometimes embedded in larger groups such as $SU(2)_L\times SU(2)_R$, $SU(4)$, $SU(5)$, and $SO(10)$. See, {\it e.g.}, refs.~\cite{Appelquist:2003uu, Appelquist:2002me}.} } 

This paper is organized as follows.
In Sec.~II, we  first set up our model. We then discuss the Higgs sector,  the active neutrinos, leptogenesis, lepton flavor violating (LFV) processes
and dark matter.
In Sec. III, we give the numerical analysis to explore the allowed parameter space of the model. 
We conclude  in Sec.~IV.


\begin{widetext}
\begin{center} 
\begin{table}[b]
\caption{Field contents of fermions 
and their charge assignments under $SU(3)_C\times SU(2)_L\times U(1)_Y\times U(1)_{B-L}$, where $i=1,2$.}
\begin{tabular}{|c||c|c|c|c|c||c|c|}\hline\hline  
Fermions& ~$Q_L$~ & ~$u_R$~ & ~$d_R$~ &~$L_L$~ & ~$e_R$~ & ~$N_{R_i}$~ & ~$N_{R_3}$~ 
\\\hline 
$SU(3)_C$ & $\bm{3}$  & $\bm{3}$  & $\bm{3}$  & $\bm{1}$  & $\bm{1}$  & $\bm{1}$  & $\bm{1}$  \\\hline 
 $SU(2)_L$ & $\bm{2}$  & $\bm{1}$  & $\bm{1}$ & $\bm{2}$ & $\bm{1}$  & $\bm{1}$ & $\bm{1}$    \\\hline 
$U(1)_Y$ & $\frac16$ & $\frac23$  & $-\frac{1}{3}$ & $-\frac12$  & $-1$ & $0$ & $0$    \\\hline
 $U(1)_{B-L}$ & $\frac13$ & $\frac13$  & $\frac13$ & $-1$  & $-1$   & $-4$    & $5$   \\\hline
\end{tabular}
\label{table-1}
\end{table}
\end{center}
\end{widetext}

\begin{table}[b]
\centering {\fontsize{10}{12}
\caption{Field contents of bosons 
and their charge assignments under $SU(3)_C\times SU(2)_L\times U(1)_Y\times U(1)_{B-L}$.}
\begin{tabular}{|c||c|c|c|c|c|}\hline\hline
  Bosons  &~ $H$~  &~ $\zeta$  ~ &~ $\eta$~ &~ $\varphi_8$~ & ~$\varphi_{10}$~ \\\hline
$SU(3)_C$ & $\bm{1}$  & $\bm{1}$  & $\bm{1}$  & $\bm{1}$ & $\bm{1}$ \\\hline 
$SU(2)_L$ & $\bm{2}$ & $\bm{2}$  & $\bm{2}$ & $\bm{1}$ & $\bm{1}$  \\\hline 
$U(1)_Y$ & $\frac12$ & $\frac12$  & $\frac12$ & $0$ & $0$    \\\hline
 $U(1)_{B-L}$ & $0$ & $-3$ & $-6$ & $8$  & $10$  \\\hline
\end{tabular}%
\label{table-2}} 
\end{table}

\section{ Model setup and phenomenologies}
First of all, we impose an additional $U(1)_{B-L}$ gauge symmetry {and add three right-handed neutral fermions $N_{R_i}(i=1,2,3)$ to} the standard model (SM),
where the right-handed neutrinos have $U(1)_{B-L}$ charges of  $-4$, $-4$ and $5$, respectively. 
Consequently, all the anomalies  to be considered are  the triangular $U(1)_{B-L}^3$ and mixed gauged-gravity $U(1)_{B-L}$ ones, which are found to be zero~\cite{Montero:2007cd, Patra:2016ofq}, due to the uniqueness of the charge assignments in the SM~\cite{GM1989}.
{We also introduce $\varphi_8$ and $\varphi_{10}$ with nonzero vacuum expectation values (VEVs) after the SSB of  $U(1)_{B-L}$. As a result,  three right-handed neutral fermions acquire nonzero Majorana masses.
Note here that one is still unable to understand active neutrino masses due to the absence of the Yukawa term $\bar L_L \tilde H N_R$.}
To solve this problem, we place $SU(2)_L$ doublet bosons $\zeta$ and $\eta$ with nonzero $U(1)_{B-L}$ charges, 
so that neutrino masses are radiatively generated at one-loop level. Here, $\zeta$ is expected to be inert, whereas $\eta$ is not.
Also the stability of DM  is assured by a remnant $Z_2$ symmetry after the SSB of $U(1)_{B-L}$.
Field contents and their assignments for fermions and bosons are  given in Table~\ref{table-1} and~\ref{table-2}, respectively.
The renormalizable Lagrangian for the lepton sector and Higgs potential are  given by 
\begin{align}
-{\cal L}_{L}&=
y_{\ell_a} \bar L_{L_a} e_{R_a} H + (y_\zeta)_{ai}\bar L_{L_a} \tilde\zeta N_{R_i}
+\frac{y_{N_{i}}}2 \bar N^C_{R_i}  N_{R_i} \varphi_8 + \frac{y_{N_3}}2 \bar N^C_{R_3}  N_{R_3} \varphi_{10}^* 
+{\rm h.c.},\label{eq:lag-lep}\\
V&=
{ {\mu_H^2} |H|^2 + {\mu_\eta^2} |\eta|^2+ {\mu_\zeta^2} |\zeta|^2 + {\mu^2_{\varphi_8}} |\varphi_8|^2 + {\mu^2_{\varphi_{10}}} |\varphi_{10}|^2} 
+\frac{\lambda_0}2\left[(H^\dag\zeta) (\eta^\dag\zeta) + {\rm h.c.}\right]\nn\\
&+
\frac{\lambda_H}4 |H|^4 + \frac{\lambda_\eta}4 |\eta|^4+ \frac{\lambda_\zeta}4 |\zeta|^4 + \frac{\lambda_{\varphi_8}}4|\varphi_8|^4
+ \frac{\lambda_{\varphi_{10}}}4|\varphi_{10}|^4  + \lambda_{H\eta} |H|^2 |\eta|^2 +\lambda'_{H\eta} |H^\dag \eta|^2 
\nn\\&  
+ \lambda_{H\zeta} |H|^2 |\zeta|^2 +\lambda'_{H\zeta} |H^\dag \zeta|^2 
+ \lambda_{H\varphi_8} |H|^2|\varphi_8|^2 + \lambda_{H\varphi_{10}} |H|^2|\varphi_{10}|^2 
+ \lambda_{\eta\zeta} |\eta|^2 |\zeta|^2 +\lambda'_{\eta\zeta} |\eta^\dag \zeta|^2
\nn\\
&
+ \lambda_{\eta\varphi_8} |\eta|^2|\varphi_8|^2 + \lambda_{\eta\varphi_{10}} |\eta|^2|\varphi_{10}|^2 
+ \lambda_{\zeta\varphi_8} |\zeta|^2|\varphi_8|^2 + \lambda_{\zeta\varphi_{10}} |\zeta|^2|\varphi_{10}|^2 
+ \lambda_{\varphi_8\varphi_{10}}|\varphi_8|^2 |\varphi_{10}|^2,
\label{eq:lag-pot}
\end{align}
respectively, 
where $\tilde \zeta \equiv (i \sigma_2) \zeta^*$ with $\sigma_2$ being the second Pauli matrix, 
and $a(i)$ runs over $1$ to $3(2)$.

In the {\it scalar sector},
the scalar fields are parameterized as 
\begin{align}
&H =\left[\begin{array}{c}
w^+\\
\frac{v + h +i z}{\sqrt2}
\end{array}\right],\;
\eta =\left[\begin{array}{c}
\eta^+\\
\frac{v_\eta + \eta_R +i \eta_I}{\sqrt2}
\end{array}\right],\;
\zeta =\left[\begin{array}{c}
\zeta^+\\
\frac{ \zeta_R +i \zeta_I}{\sqrt2}
\end{array}\right],\;
\varphi_i=
\frac{v_{\varphi_i} +\varphi_{R_i} + iz_{\varphi_i}}{\sqrt2},\ (i=8,10),
\label{component}
\end{align}
where $\sqrt{v^2+v_\eta^2}\approx$246 {GeV}, each of the lightest states ($=$massless states) of {$(w^\pm,\eta^\pm)$ and $(z, \eta_I)$, and $(z_{\varphi_{8}},z_{\varphi_{10}})$ is absorbed by the SM gauge bosons of $W^\pm$ and $Z$, and the $B-L$ gauge boson of $Z'$, induced after the SSB.
{Inserting tadpole conditions, the singly-charged mass matrix with 2 by 2 in the basis of $(w^\pm, \eta^\pm)^T$ is defined by $M_C$, which is diagonalized by the orthogonal matrix $O_C$ as $m^2_{H^\pm_i}=O_C M_C^2 O_C^T,\ (i=1,2)$ with $H^\pm_i (m_{H^\pm_i})$  the
mass eigenstates (eigenvalues), 
where
\begin{align}M_C^2&=
\frac{\lambda'_{H\eta}}2 \left[\begin{array}{cc} v_\eta^2 &  -{v v_\eta} \\ -{v v_\eta} & v^2 \\ 
\end{array}\right],\
O_C=
\left[\begin{array}{cc} \frac{v}{\sqrt{v^2+v_\eta^2}} &  \frac{v_\eta}{\sqrt{v^2+v_\eta^2}} \\ 
- \frac{v_\eta}{\sqrt{v^2+v_\eta^2}} & \frac{v}{\sqrt{v^2+v_\eta^2}} \\ 
\end{array}\right],\
m_{H^\pm_i}^2=
\left[\begin{array}{cc} 0 & 0 \\ 
0&\frac{\lambda'_{H\eta}}2(v^2+v_\eta^2) \\ 
\end{array}\right].\
\end{align}
Therefore, the structure of $(w^\pm,\eta^\pm)$ is same as one of the two Higgs doublet models~\cite{thdm}.
In the same way as the singly-charged case, 
the CP even matrix with 4 by 4 in the basis of $(h,\eta_R, \varphi_{R_8}, \varphi_{R_{10}})^T$ is defined by $M_R$, which is diagonalized by the orthogonal matrix $O_R$ as $m^2_{h_i}=O_R M_R^2 O_R^T,\ (i=1,2,\cdots,4)$  with $h_i(m_{h_i})$  the
mass eigenstates (eigenvalues). Furthermore, the SM Higgs is defined by $h_{SM}\equiv h_1$ with  
$m_{h_{SM}}\equiv m_{h_1}=125$ GeV.
The concrete form of $M_R$ is given by 
\begin{align}M_R^2&\equiv
\left[\begin{array}{cccc}
\frac{v^2\lambda_{H}}2 &  v v_\eta (\lambda_{H\eta}+\lambda'_{H\eta}) &   v v_{\varphi_8}\lambda_{H\varphi_8} &   v v_{\varphi_{10}}\lambda_{H\varphi_{10}} \\ 
v v_\eta (\lambda_{H\eta}+\lambda'_{H\eta}) &\frac{v_\eta^2\lambda_\eta}2 &  v_\eta v_{\varphi_8}\lambda_{\eta\varphi_8} &  v_\eta v_{\varphi_{10}}\lambda_{\eta\varphi_{10}}  \\ 
  v v_{\varphi_8}\lambda_{H\varphi_8}  &v_\eta v_{\varphi_8} \lambda_{\eta\varphi_8} & \frac{v_{\varphi_8}^2\lambda_{\varphi_8}}2 &  v_{\varphi_{8}} v_{\varphi_{10}}\lambda_{\varphi_{8}\varphi_{10}}  \\
v v_{\varphi_{10}}\lambda_{H\varphi_{10}}  &v_\eta v_{\varphi_{10}} \lambda_{\eta\varphi_{10}} & v_{\varphi_8}v_{\varphi_{10}}\lambda_{\varphi_8\varphi_{10}} & \frac{v_{\varphi_{10}}^2 \lambda_{\varphi_{10}}}2
  \\ \end{array}\right],\end{align}
where the $O_R$ and $m_{h_i}$ are numerically obtained.
In the inert sector, the mass eigenstates of $\zeta_{R(I)}$ and $\zeta^\pm$ are given by~\cite{Barbieri:2006dq}
\begin{align}
m_{\zeta_R}^2&= M_{\zeta}^2 + {\frac12} \lambda_0 v v_\eta,\quad
m_{\zeta_I}^2= M_{\zeta}^2 - {\frac12} \lambda_0 v v_\eta,\quad 
{ m_{\zeta^\pm}^2= M_{\zeta}^2 -\frac12 \lambda'_{\eta\zeta} v_\eta^2},\\
M_{\zeta}^2&= \mu_\zeta^2 +
{\frac12}\left[\lambda_{\zeta\varphi_8}v_{\varphi_{8}}^2 +\lambda_{\zeta\varphi_{10}}v_{\varphi_{10}}^2+(\lambda_{\eta\zeta}+\lambda'_{\eta\zeta})v_\eta^2\right].
  \end{align}
Here, we will briefly discuss the breaking scale of $U(1)_{B-L}$. 
Because of our large numbers of $B-L$ charge assignments for bosons $\varphi_8$ and $\varphi_{10}$ in Table~\ref{table-2},
our theory can really be within the TeV scale, where these bosons cause the SSB of $U(1)_{B-L}$ in order to give masses of the right-handed neutrinos with  $B-L$ charges (-4,-4,5).
The breaking scale could be evaluated by the mass of the $B-L$ gauge boson.
Once we fix the gauge coupling of $B-L$ ($g'$) to be {\cal O}(1), its typical mass is greater than 6.9 TeV from the LEP constraint~\cite{Schael:2013ita}.
On the other hand, its theoretical mass in our model can be given by $m_{Z'}=g'\sqrt{(8v_{\varphi_8})^2 +(10 v_{\varphi_{10}})^2 }$.
Even assuming $v_{\varphi_8}>>v_{\varphi_{10}}$, the typical breaking scale of $B-L$ can be less than 1 TeV.

\begin{figure}[t]
\begin{center}
\includegraphics[width=10cm]{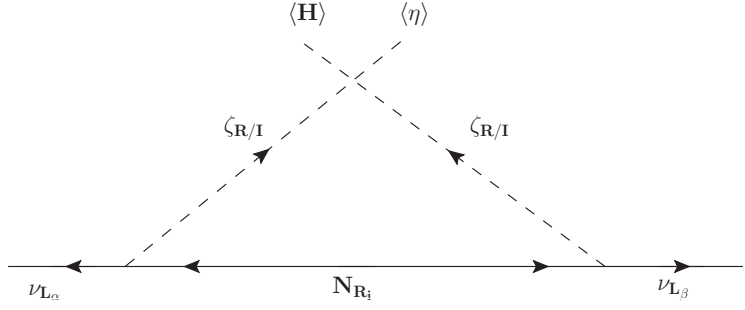} 
\caption{One loop diagram which induces neutrino masses. } 
  \label{fig:diagram}
\end{center}
\end{figure}
  \subsection{Neutrino masses}
Since $y_\ell$ and $y_{N}$ can be diagonal without loss of generality in Eq.~(\ref{eq:lag-lep}),
we define $m_{\ell_a} =y_{\ell_a} v/\sqrt2\ (a=e,\mu,\tau)$ $M_{N_i} =y_{N_i} v_{\varphi_8}/\sqrt2\ (i=1,2)$ after the electroweak
 and $U(1)_{B-L}$ symmetry breakings, where $m_\ell$ and $M_N$ are respectively the masses of charged-leptons and right-handed neutrinos.
The neutrino mass matrix is induced at  one-loop level as shown in Fig.~\ref{fig:diagram}, and its form is given by~\cite{Ma:2006km}
\be
({\cal M}_{\nu})_{\alpha\beta}  =
 \frac{1}{ 16\pi^2}
\sum_{i=1,2}  (y_\zeta)_{\alpha i}  (y_\zeta^T)_{i\beta}  M_{N_i} 
\left(\frac{m_{\zeta_R}^2}{m_{\zeta_R}^2-M^2_{N_i}}\ln\left[\frac{m_{\zeta_R}^2}{M_{N_i}^2} \right] 
-
\frac{m_{\zeta_I}^2}{m_{\zeta_I}^2-M^2_{N_i}}\ln\left[\frac{m_{\zeta_I}^2}{M_{N_i}^2} \right] 
\right).\;
\ee
 We note that the
 Casas-Ibarra parametrization is  a convenient method to achieve the numerical analysis~\cite{Casas:2001sr}.
 Once we define $m_\nu\equiv U_{MNS} {\cal M}_{\nu} U_{MNS}^T$,
 $y_\zeta$ can be replaced by observables with several arbitral parameters given by
 \bea
 y_\zeta & = & U^\dag_{MNS} m_\nu^{1/2} O R^{-1/2}\,,
 \nonumber\\
  R &\equiv &  \frac{1}{ 16\pi^2}
\sum_{i=1,2}  M_{N_i} 
\left(\frac{m_{\zeta_R}^2}{m_{\zeta_R}^2-M^2_{N_i}}\ln\left[\frac{m_{\zeta_R}^2}{M_{N_i}^2} \right] 
-
\frac{m_{\zeta_I}^2}{m_{\zeta_I}^2-M^2_{N_i}}\ln\left[\frac{m_{\zeta_I}^2}{M_{N_i}^2} \right] 
\right),
\eea
where $U_{MNS}$ and $m_\nu$ are measured by neutrino oscillation experiments.
And $O$  is an arbitral complex 3 by 2 rotation matrix
with $OO^T={\rm Diag}(0,1,1)$  ($O^TO=1_{2\times2}$), 
which can be parametrized by the following matrices for the normal hierarchy (NH) and inverted hierarchy (IH)~\cite{Rink:2016vvl}:
\begin{align}
O =\left[\begin{array}{cc}
0 & 0\\
\cos z & -\sin z \\
\pm \sin z & \pm\cos z \\
\end{array}\right], \quad 
O =\left[\begin{array}{cc}
\cos z & -\sin z \\
\pm \sin z & \pm\cos z \\
0 & 0\\
\end{array}\right],  
\label{eq:omix}
\end{align}  
respectively,
where $z$ can be complex.
In our numerical analysis, we will use the global fit of the current neutrino oscillation data as the best fit values for NH and IH~\cite{Gonzalez-Garcia:2014bfa}:
\begin{align}
{\rm NH}:\ 
&s_{12}^2=0.304,\quad s_{23}^2=0.452,\quad s_{13}^2=0.0218,\quad \delta_{CP}=\frac{306}{180}\pi,\nn\\
&\quad (m_{\nu_1},\ m_{\nu_2},\ m_{\nu_3})\approx(0, 8.66, 49.6)\ {\rm meV}, \label{eq:NH}\\ 
{\rm IH}:\ 
&s_{12}^2=0.304,\quad s_{23}^2=0.579,\quad s_{13}^2=0.0219,\quad \delta_{CP}=\frac{254}{180}\pi,\nn\\
&\quad (m_{\nu_1},\ m_{\nu_2},\ m_{\nu_3})\approx(49.5,50.2,0)\ {\rm meV},  \label{eq:IH}
\end{align}
where $s_{12,13,23}$ are the short-hand notations of $\sin\theta_{12,13,23}$ for three mixing angles of $U_{MNS}$, 
while two Majorana phases are taken to be zero.

{
\begin{figure}[t]
\begin{center}
\includegraphics[width=10cm]{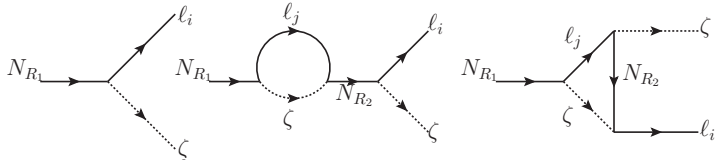} 
\caption{Tree level and one-loop diagrams for $N_{R_1}\to\ell_i\zeta$. } 
  \label{fig:leptogenesis}
\end{center}\end{figure}
\subsection{Leptogenesis}
Here, we discuss the resonant leptogenesis mechanism,  followed by those in ref.~\cite{Chen:2009gd}.\footnote{{A comprehensive study is found in, {\it e.g.}, ref.~\cite{Heeck:2016oda}.}}
First of all, we expect that the source of the CP asymmetry (CPA) is induced from $N_{R_1}$ via the two-body decay as shown in Fig.~\ref{fig:leptogenesis}.
Then, the CPA, which is denoted by $\epsilon$, is approximately computed by
\begin{align}
\epsilon\simeq -\frac{3}{16\pi} \frac{{\rm Im}\left[(y^\dag_{\zeta} y_{\zeta})^2_{12}\right]}{(y^\dag_{\zeta} y_{\zeta})_{11}} \frac{M_{N_1}}{M_{N_2}},
\label{eq:cp-vio}
\end{align}
where we have assumed $M_{N_1}<<M_{N_2}$.\footnote{In our numerical analysis, we will take $3\lesssim M_{N_1}/M_{N_2}\lesssim10$~\cite{Gu:2008yk}.}
During creating the lepton asymmetry,  the decay width of $N_{R_1}$ should satisfy the following condition of the out-of-equilibrium:
\begin{align}
&
\Gamma(N_{R_1}\to \ell^\pm \zeta^\mp)\lesssim H(M_{N_1}),
\end{align}
with 
\begin{align}
&\Gamma(N_{R_1} \to \ell^\pm \zeta^\mp)=\frac{(y^\dag_{\zeta} y_{\zeta})_{11}}{16\pi} M_{N_1}
\left(1-\frac{m^2_{\zeta^\pm}}{M_{N_1}^2}\right)^2,\quad H(T)=\left(\frac{8\pi^3g_*}{90}\right)^{1/2}\frac{T^2}{M_{Pl}},
\end{align}
where $H(T)$ is the Hubble parameter at the temperature ($T$), $g_*\approx 100$ is the relativistic degrees of freedom, and $M_{Pl}\approx10^{19}$ GeV is the Planck mass.
Furthermore, one can derive the following condition:
\begin{align}
(y^\dag_{\zeta} y_{\zeta})_{11} \lesssim 
\left(\frac{256\pi^5g_*}{45}\right)^{1/2} \frac{M_{N_1}}{M_{Pl}} \left(1-\frac{m^2_{\zeta^\pm}}{M_{N_1}^2}\right)^{-2},
\label{eq:out-eql-cond}
\end{align}
which implies $(1-m^2_{\zeta^\pm}/M_{N_1}^2) = {\cal O}(10^{-5}\sim 10^{-4})$ for $y_\zeta={\cal O}(10^{-3})$ and $M_{N_1}={\cal O}(0.1\sim 1)$ TeV, although we will numerically analyze later.
Consequently, the resulting BAU ($Y_B$) is found as~\cite{Gu:2008yk}
\begin{align}
Y_B\simeq-\frac{1}{15}\frac{\epsilon}{g_*}= (5.8\sim6.6)\times10^{-10},\label{eq:bau}
\end{align}
where the last value is the current bound on the BAU~\cite{pdg}. 

\begin{figure}[t]
\begin{center}
\includegraphics[width=10cm]{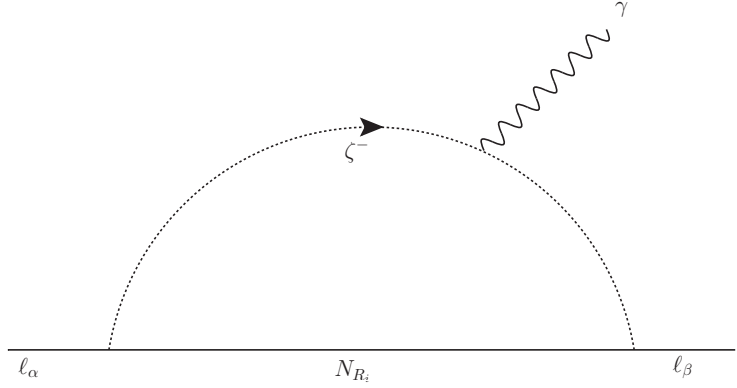} 
\caption{One loop diagram for LFVs of $\ell_\alpha\to\ell_\beta\gamma$. } 
  \label{fig:diagram1}
\end{center}\end{figure}
\begin{figure}[t]
\begin{center}
\includegraphics[width=10cm]{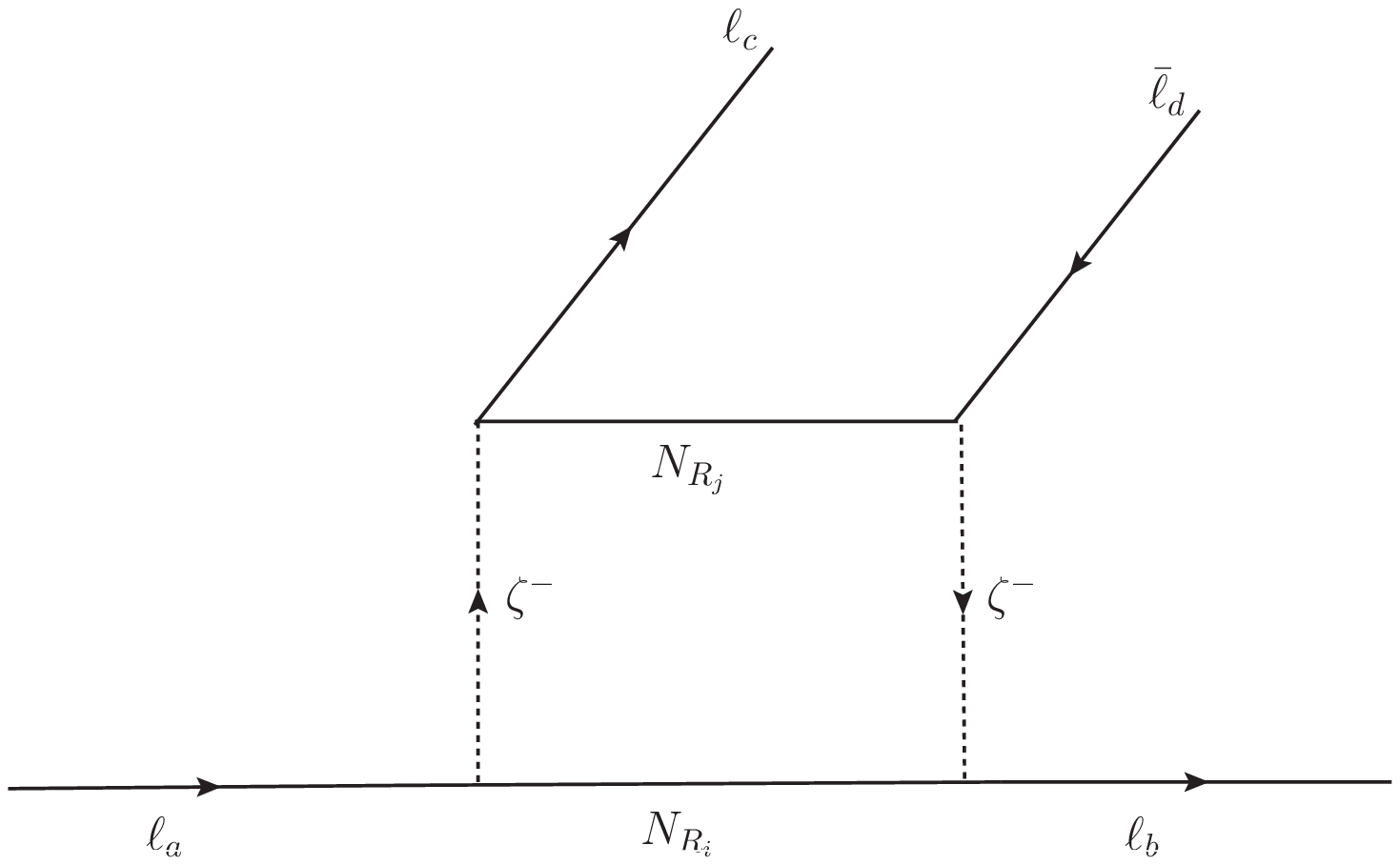} 
\caption{Typical box diagrams for LFVs of $\ell_a\to\ell_b\ell_c\bar\ell_d$.} 
  \label{fig:diagram2}
\end{center}\end{figure}
\begin{figure}[t]
\begin{center}
\includegraphics[width=7cm]{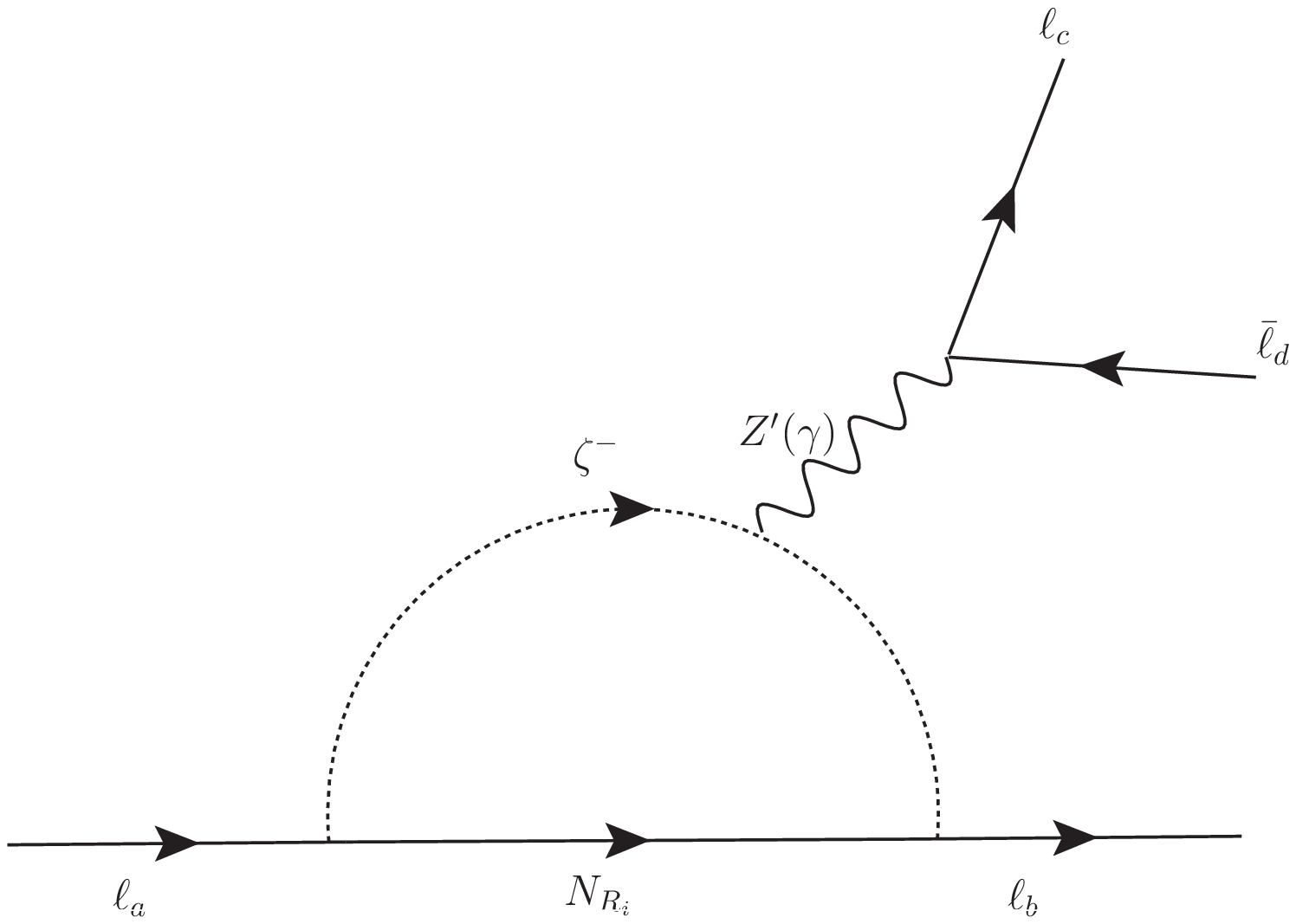} 
\includegraphics[width=7cm]{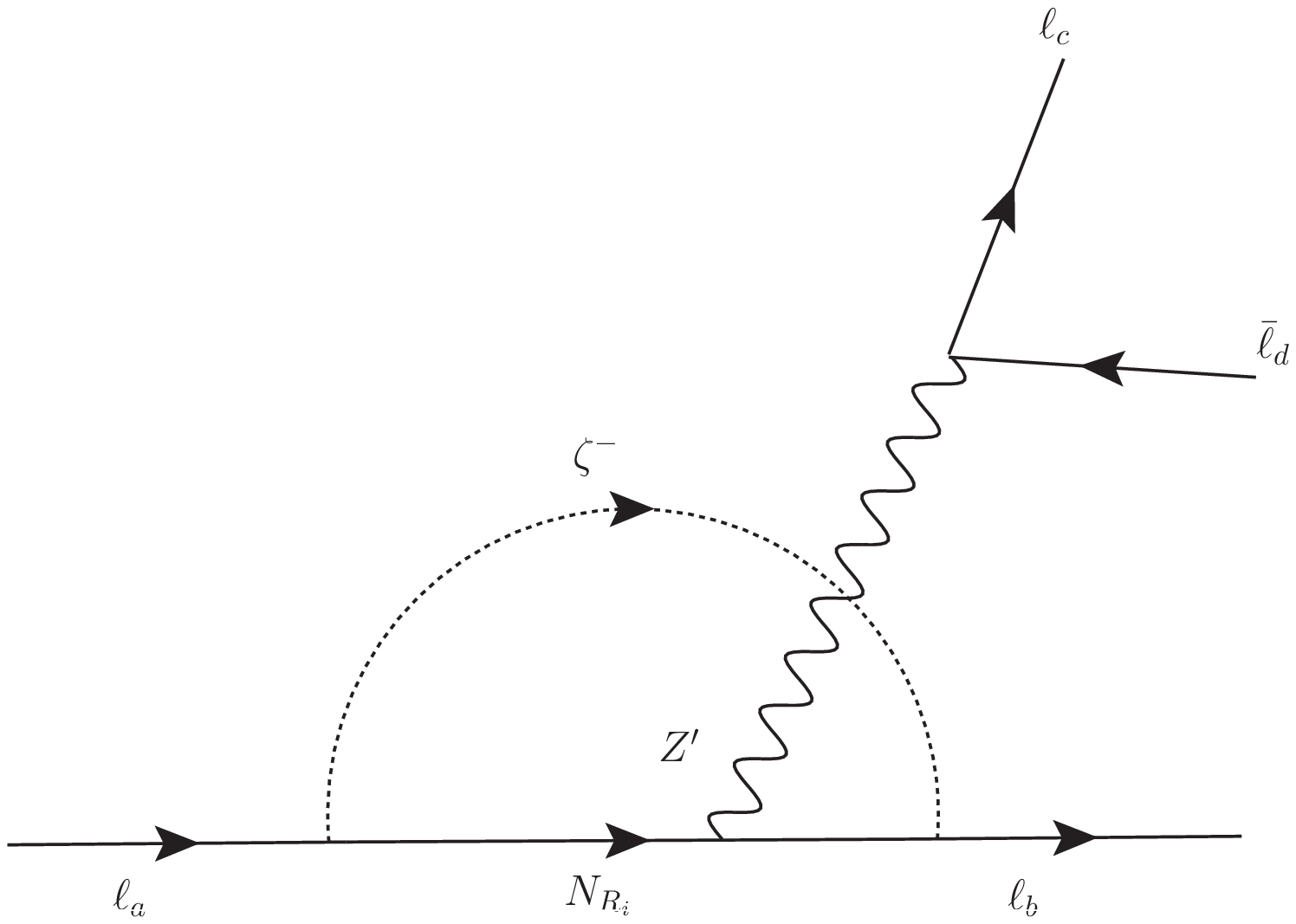} 
\caption{One-loop diagrams via the $Z'(\gamma)$ boson for LFVs of $\ell_a\to\ell_b\ell_c\bar\ell_d$.
} 
  \label{fig:diagram3}
\end{center}\end{figure}

\subsection{Lepton flavor violating processes}
In our model, there exist various LFV processes such as $\ell_\alpha\to \ell_\beta\gamma$, $\ell_a\to\ell_b\ell_c\bar\ell_d$,
and flavor changing processes involving quarks such as semi-leptonic decays.
First of all, let us consider
the processes $\ell_\alpha \to \ell_\beta \gamma$ in Fig.~\ref{fig:diagram1}, which
are induced from the neutrino Yukawa couplings at one-loop level. The decay branching ratios (BRs) are given by
\begin{align}
{\rm BR}(\ell_\alpha\to \ell_\beta \gamma)&=\frac{3\alpha_{em}C_{\alpha\beta}}{16\pi G_F^2}
\left|\sum_{i=1,2}\frac{(y_\zeta)_{\beta i} (y_\zeta^\dag)_{i\alpha}}{M^2_{N_i}} F_I( r_{\ell_\alpha}, r_{N_i})\right|^2,\\
F_I(r_{\ell_\alpha}, r_{N_i})&= \int_0^1 dx\int_0^{1-x}dy
\frac{x y } {x+(1-x) r_{N_i} +(x^2-x) r_{\ell_\alpha} }, \label{eq:fi}
%
\end{align}
where $r_{N_i}\equiv m_{\zeta^\pm}^2/M_{N_i}^2$, $r_{\ell_\alpha}\equiv m_{\ell_\alpha}^2/M_{N_i}^2$, $\alpha_{em}\approx1/137$, $G_F\approx1.17\times10^{-5}$ GeV$^{-2}$,
$C_{\mu e}\approx1$, $C_{\tau e}\approx 0.1784$, and $C_{\tau\mu}\approx0.1736$. 
The simplified form in the limit of $r_{\ell_\alpha}<<1$ is read as
\begin{align}
 F_I(0,r_{N_i})\approx \frac{2 +3 r_{N_i} -6 r_{N_i}^2+ r_{N_i}^3 +6 r_{N_i} \ln r_{N_i} } {(1-r_{N_i})^4 }.
\end{align}
Experimental upper bounds for the branching ratios of the LFV decays are found to be~\cite{TheMEG:2016wtm}: 
\begin{equation}
{\rm BR}(\mu\to e \gamma)\lesssim 4.2\times 10^{-13},\ 
{\rm BR}(\tau\to e \gamma)\lesssim 3.3\times 10^{-8},\ 
{\rm BR}(\tau\to \mu \gamma)\lesssim 4.4\times 10^{-13}.
 \end{equation}
The stringent constraint comes from $\mu\to e\gamma$, and it roughly gives the upper limit $y_\zeta\lesssim {\cal O}(10^{-2})$
for $M_{N_1}={\cal O}(100)$ GeV, where $M_{N_1}<<M_{N_2}$ is expected from the leptogenesis.

\begin{table}[t]
\caption{A summary of the constraints of three-body decay branching ratios (BRs).}
\begin{tabular}{|c|} \hline
{Upper limits} of BRs  \\ \hline
~~~${\rm BR}(\mu^-\to e^+e^-e^-)\lesssim 1.0\times 10^{-12}$~~~ \\
~~~${\rm BR}(\tau^-\to e^+e^-e^-) \lesssim2.7\times 10^{-8}$~~~  \\
~~~${\rm BR}(\tau^-\to e^+e^-\mu^-) \lesssim1.8\times 10^{-8}$~~~ \\
~~~${\rm BR}(\tau^-\to e^+\mu^-\mu^-) \lesssim1.7\times 10^{-8}$~~~ \\
~~~${\rm BR}(\tau^-\to \mu^+e^-e^-) \lesssim1.5\times 10^{-8}$~~~  \\
~~~${\rm BR}(\tau^-\to \mu^+e^-\mu^-) \lesssim2.7\times 10^{-8}$~~~ \\
~~~${\rm BR}(\tau^-\to \mu^+\mu^-\mu^-) \lesssim2.1\times 10^{-8}$~~~  \\ \hline
\end{tabular}
\label{tab:LFVconst1}
\end{table}

Next, let us discuss the three-body LFV decays of $\ell_a\to\ell_b\ell_c\bar\ell_d$.
These decays are of course expected to be tiny compared to those of $\ell_\alpha\to\ell_\beta\gamma$ due to phase spaces as well as additional small couplings. In our case, they consist of two kinds of one-loop diagrams; box diagrams via Yukawa couplings in Fig.~\ref{fig:diagram2} and Penguin types of diagrams mediated by $\gamma(Z')$ boson via kinetic and Yukawa terms in Fig.~\ref{fig:diagram3}.
However since 
the diagram via $\gamma$ is 
always smaller than the contributions of $\ell_\alpha\to\ell_\beta\gamma$~\cite{Toma:2013zsa},
it can be negligible.
From the box diagrams in Fig.~\ref{fig:diagram2}, the most stringent bound for the three-body decays arises from $\mu\to ee\bar e$ as shown in Table~\ref{tab:LFVconst1}, given by~\cite{pdg}
\begin{align}
{\rm BR}(\mu\to ee\bar e)&\lesssim 1.0\times 10^{-12}.\label{eq:meee}
\end{align}
Here, we will estimate the theoretical bound in terms of the Yukawa coupling, by applying this experimental bound in Eq.~(\ref{eq:meee}).
The bound on the Yukawa coupling for the box diagram in Fig.~\ref{fig:diagram2} is severely estimated as~\cite{Crivellin:2013hpa}
\begin{align}
(y_\zeta)_{11} (y_\zeta)_{11}^\dag\lesssim 8.6\times10^{-5}\frac{M_{N_1}}{\rm GeV}.
\end{align}
Even when we take $M_{N_1}={\cal O}(100)$ GeV that is the minimal mass allowed by leptogenesis, 
we find $(y_\zeta)_{11} (y_\zeta)_{11}^\dag\lesssim{\cal O}(0.01)$; ${\rm Min.}[y_\zeta]\simeq{\cal O}(0.1)$.
This bound is weaker than the case of $\ell_\alpha\to\ell_\beta\gamma$ by one order of magnitude.
From the $Z'$ mediated diagrams in Fig.~\ref{fig:diagram3}, one finds~\cite{Crivellin:2013hpa}
\begin{align}
(y_\zeta)_{11} (y_\zeta)_{11}^\dag\lesssim 2.62\times10^{-9} \left[\frac{m_{Z'}}{{\rm GeV}}\right]^2 g'^{-2}\lesssim 0.13,
\label{eq:y11cond}
\end{align}
where we have applied the relation $(g'/m_{Z'})^2<(6.9\ {\rm TeV})^{-2}$
given by the LEP experiment~\cite{Schael:2013ita}.
Clearly, Eq.~(\ref{eq:y11cond}) leads to ${\rm Min.}[y_\zeta]\simeq{\cal O}(0.4)$, which is weaker than the box one.

 Semi-leptonic decays also occur at one-loop level via the $Z'$ boson in Fig.~\ref{fig:semi-lept},
where we neglect the contribution from the Yukawa coupling because it is tiny enough. 
The effective Hamiltonian is given by
\begin{align}
{\cal H}_{\rm eff}\approx -g_2^2 \sum_{i=u,c,t}\frac{(V^\dag_{\rm CKM})_{ki}(V_{\rm CKM})_{ij} F(u_i,W)}{(4\pi)^2}
\left[\frac{g'}{m_{Z'}}\right]^2(\bar d_k\gamma_aP_Ld_j)(\bar\ell\gamma^a\ell),
\end{align}
where $g_2$ is the $SU(2)_L$ gauge coupling, $V_{\rm CKM}$ is the Cabibbo-Kobayashi-Maskawa mixing matrix,
and $F(u_i,W)$ is an one-loop function that is order one at most.
One of the stringent constraints comes from BR$(B_s\to \bar\mu\mu)$, and it is evaluated by the coefficient of the above effective Hamiltonian, given by
 \begin{align}
{\rm BR}(B_s\to\bar\mu\mu): 
\left|g_2^2 \sum_{i=u,c,t}\frac{(V^\dag_{\rm CKM})_{si}(V_{\rm CKM})_{ib}}{(4\pi)^2}
\left[\frac{g'}{m_{Z'}}\right]^2\right|\lesssim 5\times10^{-9}\ {\rm GeV}^{-2},
\label{BRBs}
\end{align}
where we have assumed $F(u_i,W)\approx1$ and the right-hand side is the experimental bound of ${\rm BR}(B_s\to\bar\mu\mu)$.
As the numerical value of the left-hand side in Eq.~(\ref{BRBs}) is the order $10^{-12}$ at most by applying the relation $(g'/m_{Z'})^2<(6.9\ {\rm TeV})^{-2}$, 
it is obviously within the experimental result.\footnote{See ref.~\cite{Carpentier:2010ue} for the other experimental bounds of semi-leptonic decays.}
Thus, we will only consider the processes of $\ell_\alpha\to\ell_\beta\gamma$ in our numerical analysis below.

\begin{figure}[t]
\begin{center}
\includegraphics[width=10cm]{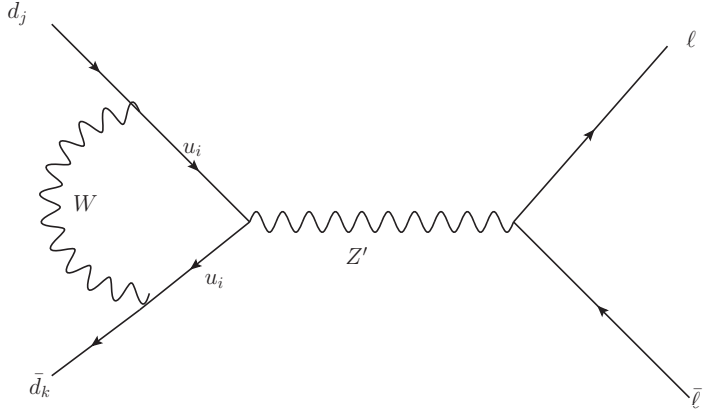} 
\caption{Typical semi-leptonic decay mode. } 
  \label{fig:semi-lept}
\end{center}\end{figure}

\begin{figure}[t]
\begin{center}
\includegraphics[width=10cm]{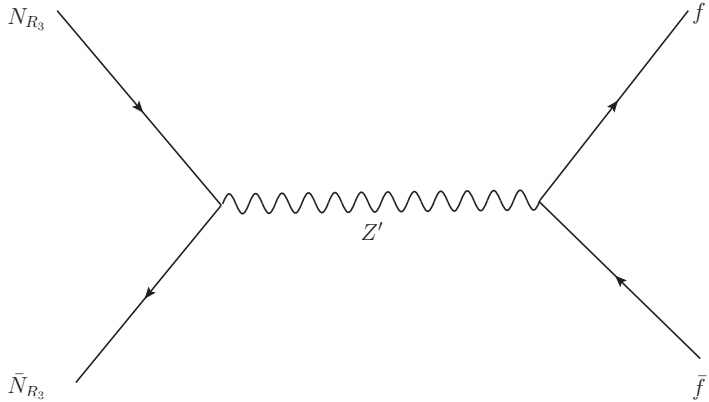} 
\caption{Main contribution to induce the relic density of DM. } 
  \label{fig:dm-annihi}
\end{center}\end{figure}
\subsection{Dark matter}
In our model,  the possible DM candidates are $\zeta_{R/I}$ and the lightest one in $N_{R_{1,2,3}}$.
For the scalar DM of $\zeta_{R/I}$,
its nature is similar to the isospin doublet inert boson~\cite{Hambye:2009pw} except Yukawa and additional (gauged) boson interactions.
However, since our typical scale of the Yukawa coupling $y_\zeta$ is around $10^{-3}$, its modes cannot be dominant to explain the relic density of DM.
Clearly, $\zeta_{R/I}$ cannot help to rely on the interactions of the Higgs potential and/or kinetic term.
Nevertheless,  it is known that there exist a lot of solutions 
to satisfy the present exclusion limits from the direct  DM experimental detections, even when a model is minimal.
To escape the limits from the spin independent direct detection searches reported by  LUX~\cite{Akerib:2016vxi}, XENON1T~\cite{Aprile:2017iyp}, and PandaX-II~\cite{PandaX},
we have to consider two dominant modes from $Z^{(')}$ and CP-even Higgs portals.
The former one with the $Z$ boson mediation can easily be evaded by giving the mass difference between $\zeta_{R}$ and $\zeta_{I}$ 
to be greater than {\cal O}(100) keV. Note  that the constraint with the $Z'$ mediation is always weaker than the $Z$ one, 
since its cross section is proportional to $(g'/m_{Z'})^2<(6.9\ {\rm TeV})^{-2}$.
%
The latter one can also be used to avoid the limits by taking  the corresponding quartic Higgs couplings, which can be written in terms of linear combinations $\lambda_0,\ \lambda_{H\zeta}^{(')},\ \lambda_{\eta\zeta}^{(')},\ \lambda_{\zeta\varphi_{8(10)}}$, to be less than 0.01, when the SM Higgs is mediating and the other masses of CP-even Higgses are assumed to be heavier than the mass of the SM Higgs.
{As a summary of the bosonic DM candidate, one finds the solution at around $500$ GeV of DM to satisfy the relic density of DM through the kinetic term.
This is almost the same as the result of the DM model with one inert two-Higgs doublet~\cite{Hambye:2009pw}.}

In case of the fermion DM candidate such as  the lightest state of $N_{R_{1,2,3}}$,
its main modes to the relic density can be found in the kinetic term with the additional gauge boson, and/or the Higgs potential.
As for both modes, its solution tends to be at around the pole with half masses of mediating fields.
 A comprehensive analysis has recently been done in refs.~\cite{Singirala:2017see, Singirala:2017cch, Nomura:2017vzp}. 
To satisfy the direct detection bounds, we have the similar processes from $Z^{'}$  and CP-even Higgses portals.
With the same reason as the $Z'$ mediation, there is almost no constraint.
As a result, we should consider the CP-even Higgses only.
The exclusion limits can be evaded by taking  the corresponding quartic Higgs couplings of  $y_{N_i,3}<$0.01~\cite{Kanemura:2010sh}. 
Once we identify either of $N_{R_{1,2}}$ as DM, the allowed parameter space might be restricted a little~\footnote{Since the related Yukawa coupling $y_\zeta$ cannot be order one in order to lead the successful resonant leptogenesis, our final result does not change drastically.}.
On the other hand, if $N_{R_{3}}$ is DM, we can discuss the DM issue independently.
{Once the leptogensis effect is taken into consideration, $N_{R_{1,2}}$ cannot be a DM candidate because of their appropriate decays.
As a result, the reasonable DM candidate is $N_{R_3}$, and its dominant contribution is expected to be in the s-channel via the $Z'$ boson in Fig.~\ref{fig:dm-annihi}.
The relic density of DM is formulated by~\cite{Edsjo:1997bg}
\begin{align}
&\Omega h^2
\approx 
\frac{1.07\times10^9}{\sqrt{g_*(x_f)}M_{Pl} J(x_f)[{\rm GeV}]},
\label{eq:relic-deff}
\end{align}
where $x_f$ is assumed to be around $25$,
and $J(x_f) (\equiv \int_{x_f}^\infty dx \frac{\langle \sigma v_{\rm rel}\rangle}{x^2})$ is evaluated by
\begin{align}
J(x_f)&=\int_{x_f}^\infty dx\left[ \frac{\int_{4 M_{N_3}^2}^\infty ds\sqrt{s-4 M_{N_3}^2} W(s) K_1\left(\frac{\sqrt{s}}{M_{N_3}} x\right)}{16  M_{N_3}^5 x [K_2(x)]^2}\right],\\ 
W(s)
&\approx \frac{25 g'^4  (s - M_{N_3}^2)}{72\pi |s-m_{Z'}^2+i m_{Z'} \Gamma_{Z'}|^2}
\left(2\sqrt{1-\frac{4 m_t^2}{s}} (2m_t^2+s)+131 s\right),\\
\Gamma_{Z'}&\approx \frac{13 g'^2 m_{Z'}}{24\pi},
\label{eq:relic-deff}
\end{align}
with $m_t$ the top quark mass.
In Fig.~\ref{fig:relic}, we show the relic density  of DM as a function of $M_{N_3}$, where red(blue) line corresponds to $m_{Z'} = 350(700)$ GeV with $g' = 0.05$. As a trivial result,
we find that the correct relic density can be obtained near the half-mass of $Z'$ for each benchmark point.

\begin{figure}[t]
\centering
\includegraphics[width=10cm]{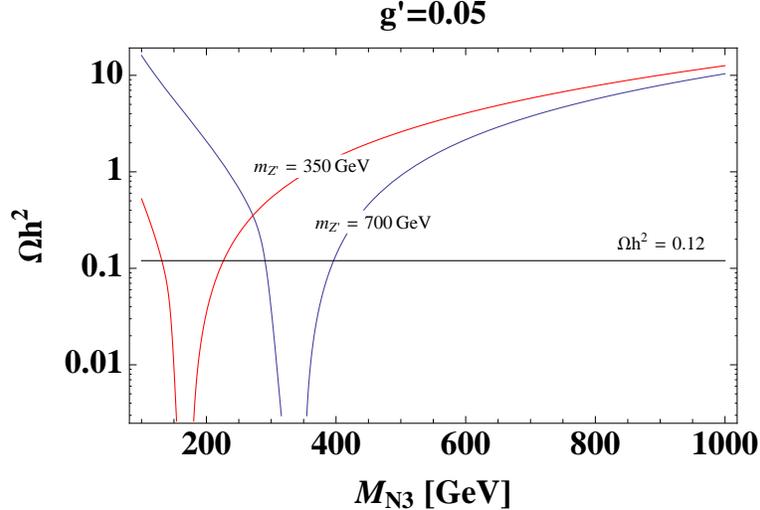}
\caption{The relic density of dark matter as a function of $M_{N_3}$, where red(blue) line corresponds to $m_{Z'} = 350(700)$ GeV with $g' = 0.05$.}
\label{fig:relic}
\end{figure}

}

\section{Numerical analysis}
For the numerical calculations, we use the neutrino oscillation data in Eqs.~(\ref{eq:NH}) and (\ref{eq:IH})
as well as the following input parameters:{
\begin{align}
& {\rm Re}[z]\in (0,\pi),\quad {\rm Im}[z]\in (-10,-1),\\
& (m_{\zeta_R}, M_{N_1}) \in (0.1,10)\ {\rm TeV},\quad M_{N_2}\in (3 \times M_{N_1},10 \times M_{N_1})\ {\rm TeV} ,
\end{align}
where we have taken $m_{\zeta_I}=m_{\zeta^\pm}$  to avoid  the constraints from
the oblique parameters~\cite{Barbieri:2006dq}, and
 $3\lesssim M_{N_1}/M_{N_2}\lesssim10$~\cite{Gu:2008yk} 
because the number density of $N_{R_2}$ immediately decreases at the temperature below $M_{N_{2}}$.}

{
Imposing the neutrino oscillation data, constraints from LFVs, and the BAU in~Eq.~(\ref{eq:bau}) with the out-of-equilibrium condition in Eq.~(\ref{eq:out-eql-cond}), we present our numerical analysis below.
First of all, we show the benchmark point for several important values in the case of NH and IH in Table~\ref{tab:bench-mark} to easily confirm they satisfy experimental results as discussed above.
In Fig.~\ref{fig:mn1-rN1}, we estimate the important value of $1-{m_{\zeta^+}^2}/{M_{N_1}^2}$ to obtain a sizable BAU in terms of $M_{N_1}$ for the cases of NH(left-side) and IH(right-side).
When $M_{N_1}\lesssim$ 1 TeV, $1-{m_{\zeta^+}^2}/{M_{N_1}^2} \lesssim 10^{-5}$ is required. Otherwise, the degeneracy between $m_{\zeta^+}$ and $M_{N_1}$ becomes to be milder slightly.
In Fig.~\ref{fig:mn1-meg}, we show the scattering plots in the plane of BR$(\mu\to e\gamma)$ and $M_{N_1}$ for NH and IH.
We see that all values for whole the range of $M_{N_1}$ are below the current experimental bound.
This is the trivial consequence in order to get the sizable BAU via leptogenesis.
In Fig.~\ref{fig:yukawas-y11-12}, we display the plots of $y_{\zeta_{11}}-y_{\zeta_{12}}$ for NH and IH, where we separate $y_{\zeta_{11}}-y_{\zeta_{12}}$ into the real and imaginary parts.
These figures suggest that the IH case is greater than the NH one by a few times.
%
In Fig.~\ref{fig:yukawas-y31-32}, we illustrate the plots of $y_{\zeta_{31}}-y_{\zeta_{32}}$ similar to Fig.~\ref{fig:yukawas-y11-12}.
These figures indicate that the NH case is slightly greater than the IH one, opposite to Fig.~\ref{fig:yukawas-y11-12}.
}

\begin{center}
\begin{table}[t]
\caption{Bench mark points (BPs) for several representative parameters  of NH and IH.}
\label{tab:bench-mark}
\begin{tabular}{c||c|c|c|c|c|c|c|c|c|c}\hline\hline  
 &$z$ & $(y^\dag_{\zeta}y_\zeta)_{11}$  & $Y_B$ & ${\rm Br}(\mu\to e\gamma)$ &  ${\rm Br}(\tau\to e\gamma)$ & ${\rm Br}(\tau\to \mu\gamma)$ & $\frac{m_{\zeta_R}}{\rm TeV}$ & $\frac{m_{\zeta^\pm}}{\rm TeV}$ & $\frac{M_{N_1}}{\rm TeV}$ &~$\frac{M_{N_2}}{\rm TeV}$\\\hline 
NH 
& $2.8$-$8.2i$ & $2.7\times10^{-5}$ & $6.0\times10^{-10}$ &$6.2\times10^{-20}$
& $1.0\times10^{-22}$
& $4.7\times 10^{-22}$ & $0.20$ & $1.3$  & $1.3$  & $7.0$ \\\hline 
IH  
& $1.4$-$9.0i$ & $3.2\times10^{-5}$ & $6.0\times10^{-10}$ &$3.1\times10^{-22}$
& $2.2\times10^{-25}$
& $1.5\times 10^{-25}$ & $0.81$ & $7.7$  & $7.7$  & $6.4$  \\\hline
\end{tabular}
\end{table}
\end{center}

\begin{figure}[t]
\begin{center}
\includegraphics[width=70mm]{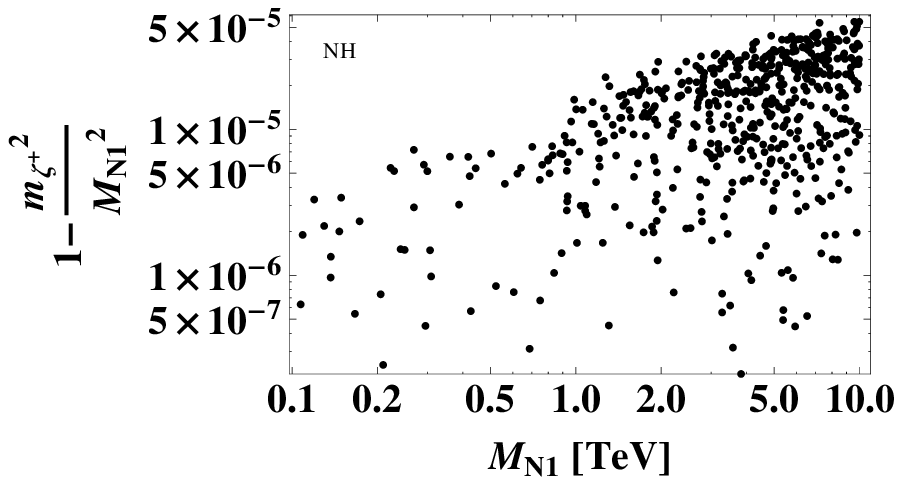} 
\includegraphics[width=70mm]{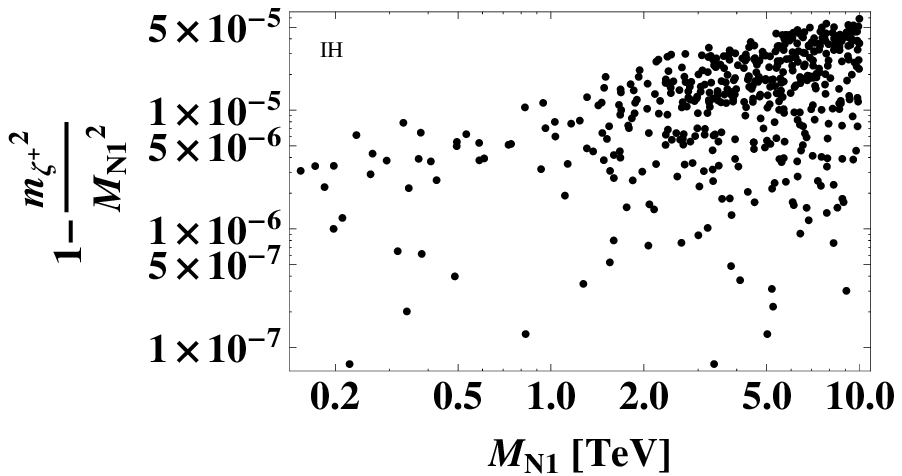} 
\caption{Scattering plots to satisfy all the data in the plane of $M_{N_1}$ and $1-{m_{\zeta^+}^2}/{M_{N_1}^2}$, where the left and right
figures correspond to NH and IH, respectively. } 
  \label{fig:mn1-rN1}
\end{center}\end{figure}

\begin{figure}[t]
\begin{center}
\includegraphics[width=70mm]{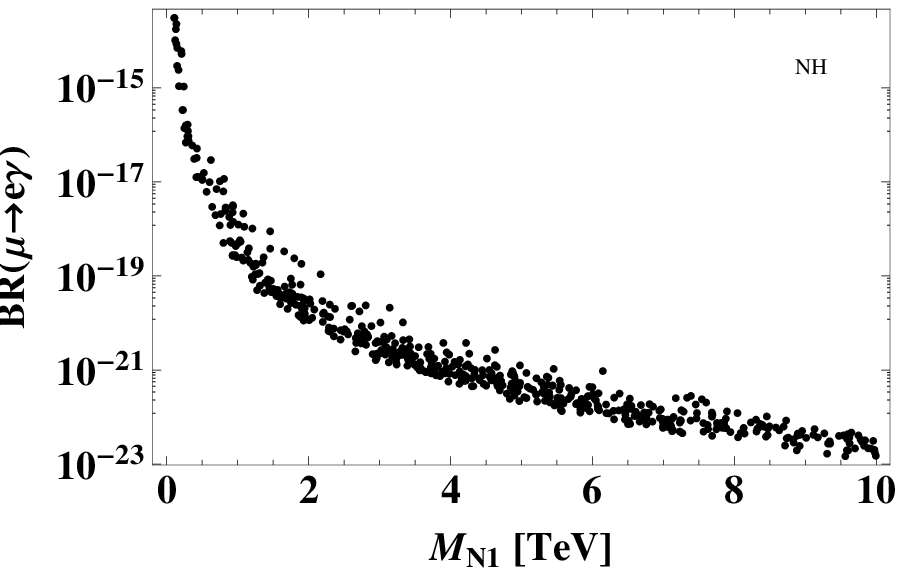} 
\includegraphics[width=70mm]{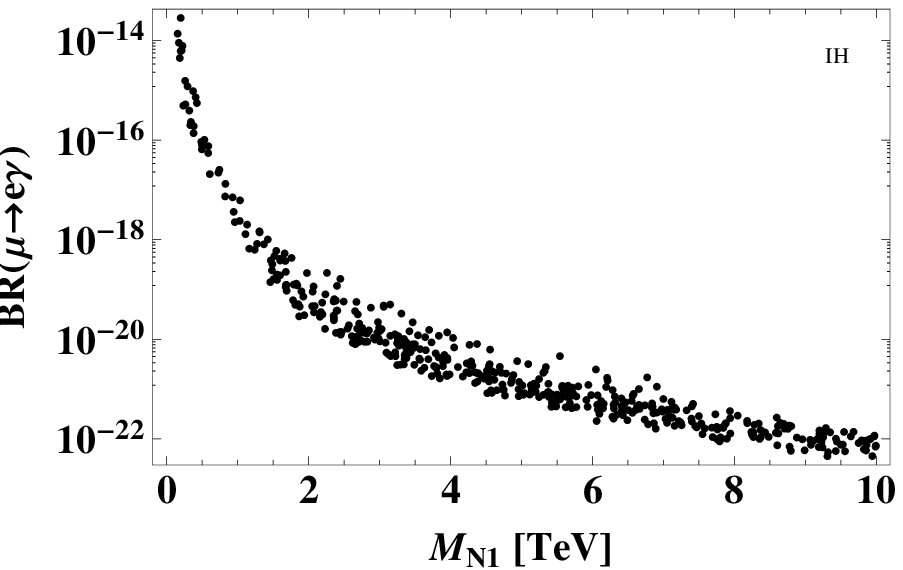} 
\caption{Scattering plots to satisfy all the data in the plane of BR$(\mu\to e\gamma)$ and $M_{N_1}$, where the left and right
figures correspond to NH and IH, respectively. } 
  \label{fig:mn1-meg}
\end{center}\end{figure}

\begin{figure}[t]
\begin{center}
\includegraphics[width=70mm]{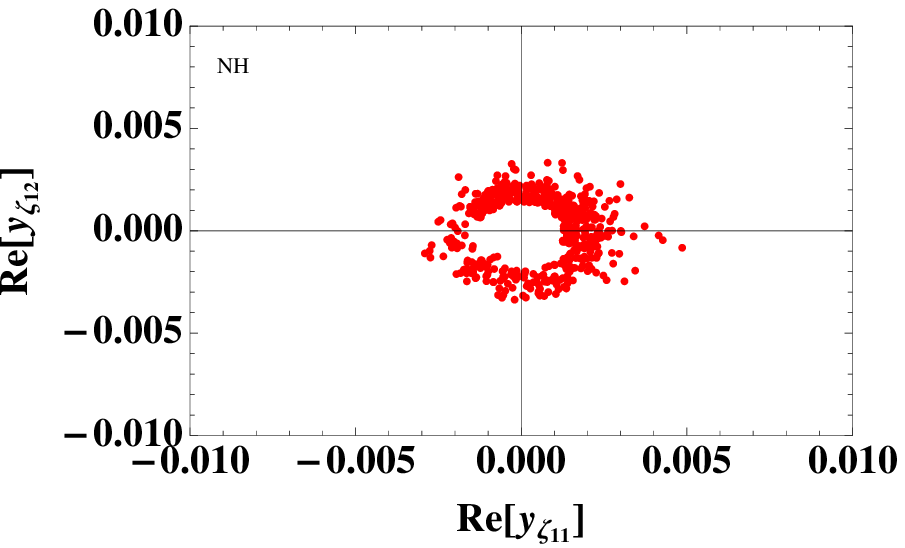} 
\includegraphics[width=70mm]{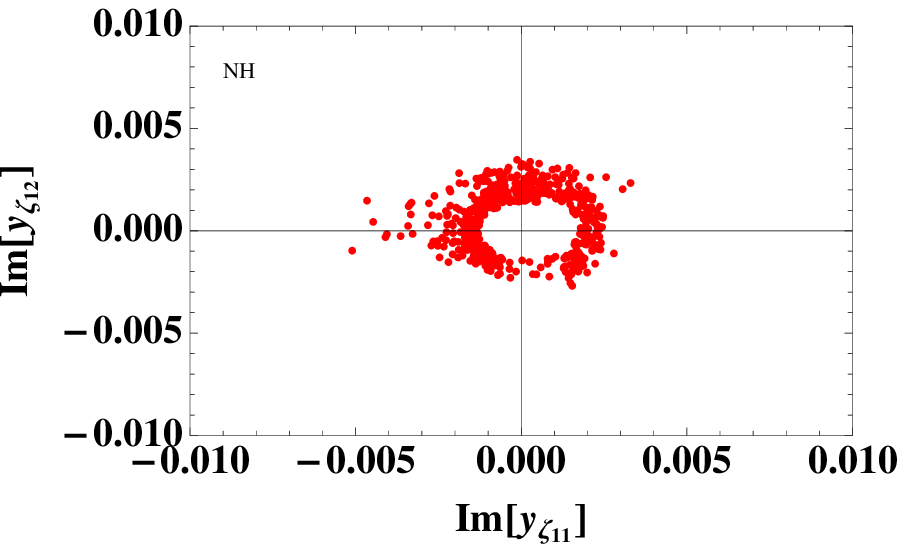} \\
\includegraphics[width=70mm]{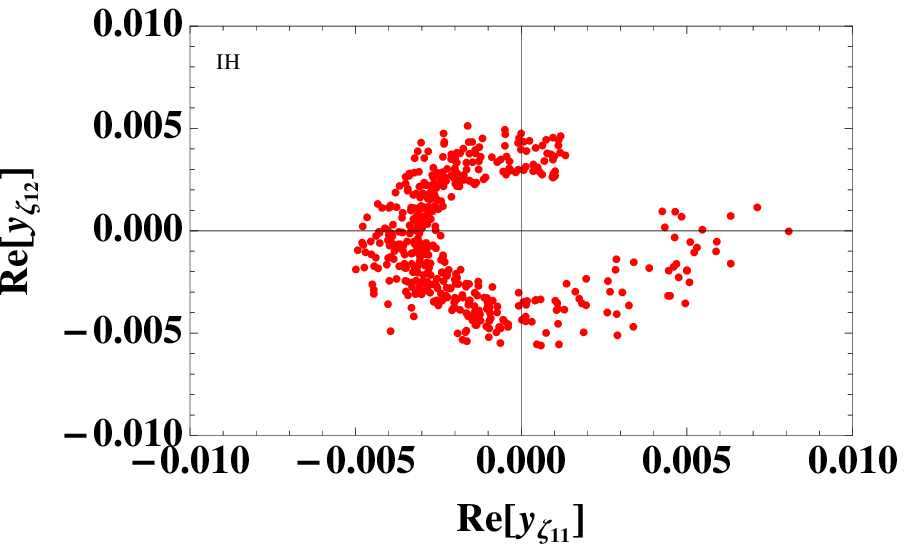} 
\includegraphics[width=70mm]{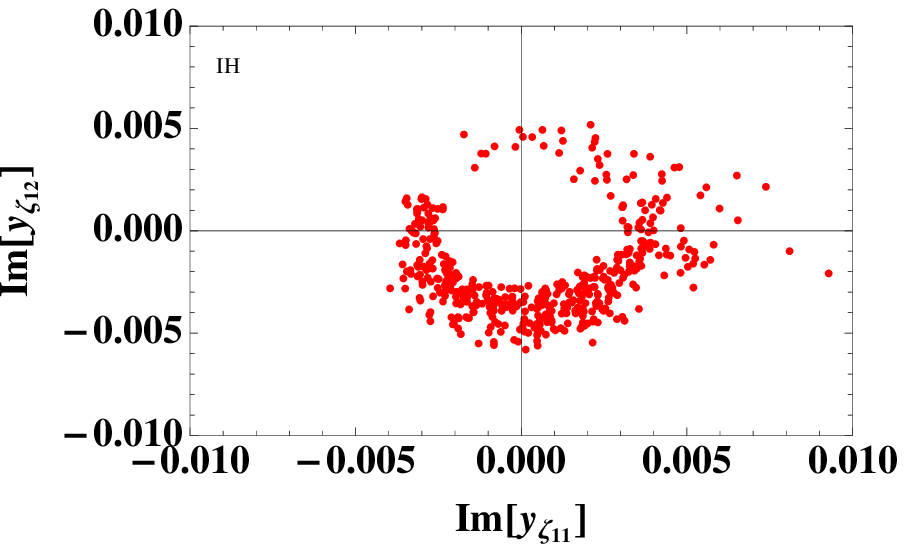} \\
\caption{Scattering plots to satisfy all the data in the planes of  ${\rm Re}[y_{\zeta_{11}}]-{\rm Re}[y_{\zeta_{12}}]$ (left-side)
and  ${\rm Im}[y_{\zeta_{11}}]-{\rm Im}[y_{\zeta_{12}}]$ (right-side), where the top (bottom) figure represents  NH (IH). } 
  \label{fig:yukawas-y11-12}
\end{center}\end{figure}

\begin{figure}[t]
\begin{center}
\includegraphics[width=70mm]{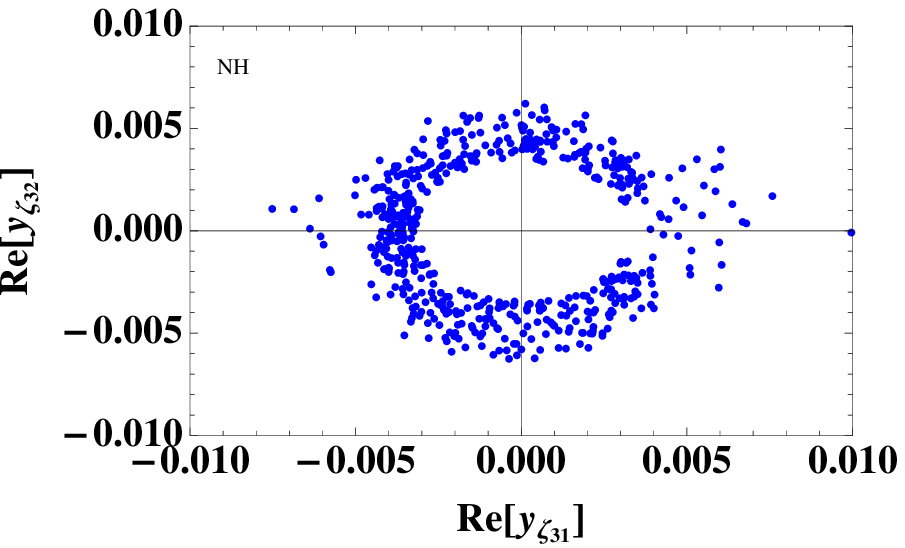} 
\includegraphics[width=70mm]{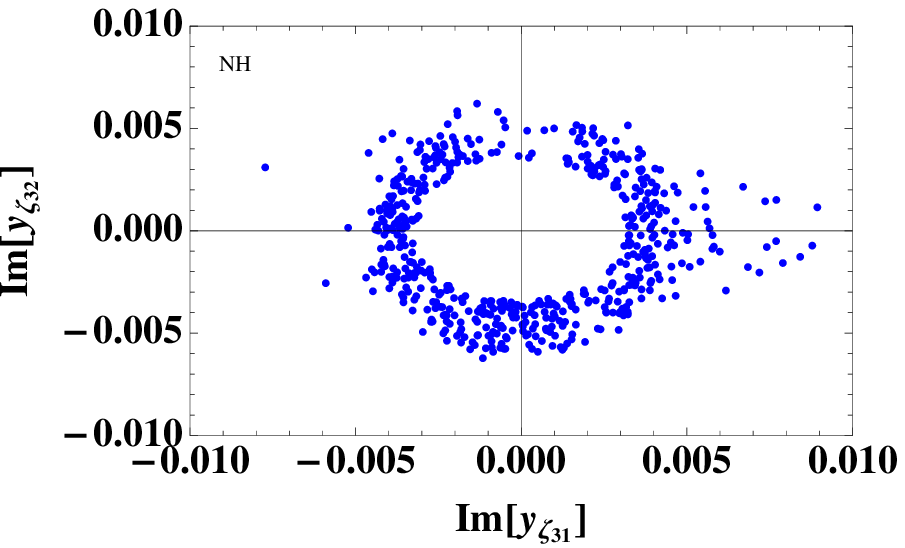} \\
\includegraphics[width=70mm]{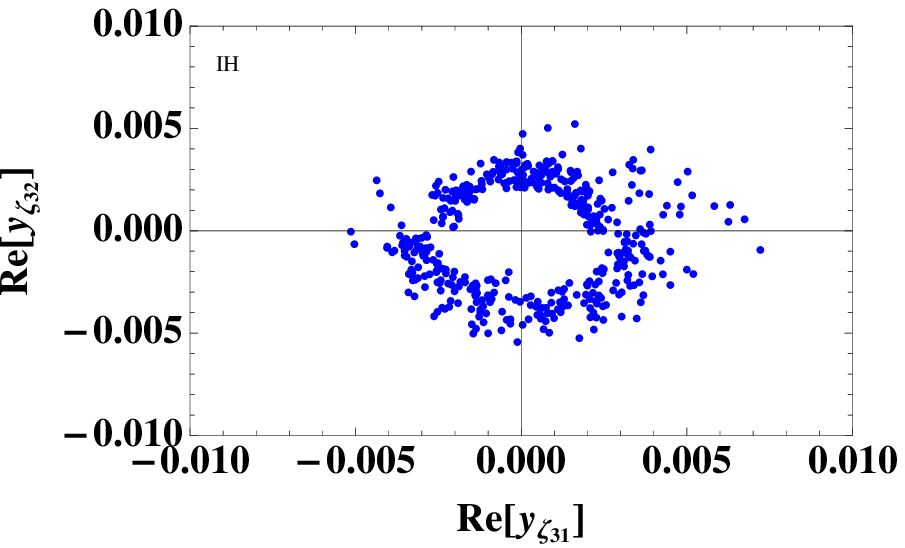} 
\includegraphics[width=70mm]{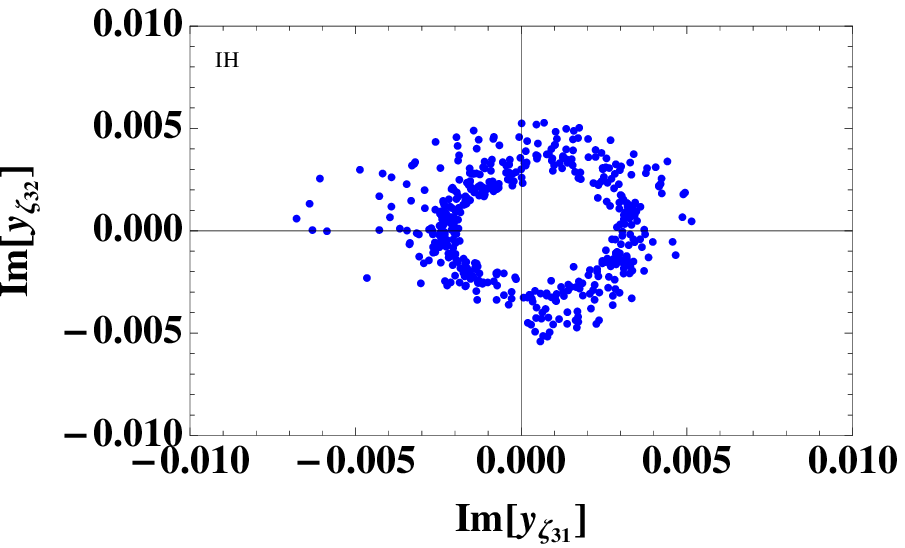} \\
\caption{Scattering plots to satisfy all the data in the planes of  ${\rm Re}[y_{\zeta_{31}}]-{\rm Re}[y_{\zeta_{32}}]$ (left-side)
and  ${\rm Im}[y_{\zeta_{31}}]-{\rm Im}[y_{\zeta_{32}}]$ (right-side), where the top (bottom) figure represents  NH (IH). } 
  \label{fig:yukawas-y31-32}
\end{center}\end{figure}

\section{Conclusion}
We have proposed a radiatively generated neutrino mass model with a successful leptogenesis to produce the BAU at TeV scale,
which contains a gauged $U(1)_{B-L}$ symmetry with unusual charge assignments of (-4,-4,5) to the right-handed neutrinos.
In this model, DM candidates naturally arrive without imposing 
any additional symmetry  to stabilize DM, which is achieved  by the resulting symmetry after the SSB of 
  $U(1)_{B-L}$.
  We have  examined the allowed regions for the model parameters  to satisfy all the experimental  constraints.
  We have found that the Yukawa couplings $y_{\zeta_{\alpha\beta}}$ with a typical order of $10^{-3}$ lead to small
  LFV processes, which cannot be measured at the current experiments.

\section*{Acknowledgments}
This work was supported in part by National Center for Theoretical Sciences and
MoST (MoST-104-2112-M-007-003-MY3).

\end{document}